\documentclass[useAMS,usenatbib]{mn2e}
\usepackage[]{graphicx}

\title[Accretion disk in the eclipsing binary AU Mon]{Accretion disk in the eclipsing binary AU Mon
\thanks{Based on photometry collected by the CoRoT space mission. The CoRoT space mission
was developed and is operated by the French space agency CNES, with participation of ESA's
RSSD and Science Programmes, Austria, Belgium, Brazil, Germany, and Spain.}}
\author[G. Djura\v sevi\'c et al.]{G. Djura\v sevi\'c$^{1,2}$\thanks{E-mail: gdjurasevic@aob.rs}, O. Latkovi\'c$^{1}$, I. Vince$^{1}$ and A. Cs\'eki$^{1}$ \\
$^{1}$Astronomical Observatory, Volgina 7, 11060 Belgrade, Serbia\\
$^{2}$Isaac Newton Institute of Chile, Yugoslavia Branch\\}

\begin{document}

\date{Accepted 2010 July 6.  Received 2010 July 4; in original form 2010 April 12}

\pagerange{\pageref{firstpage}--\pageref{lastpage}} \pubyear{2010}

\maketitle

\label{firstpage}

\begin{abstract}
We analyze the CoRoT and V-passband ground-based light curves of the interacting close binary AU Mon, assuming that there is  a geometrically and optically thick accretion disk around the hotter and more massive star, as inferred from photometric and spectroscopic characteristics of the binary. Our model fits the observations very well and provides estimates for the orbital elements and physical parameters of the components and of the accretion disk.
\end{abstract}


\section{Introduction}

AU Mon (HD 50846, HIP 33237) is an eclipsing, double-lined spectroscopic binary. The system consists of a Be-type primary (which will be referred to as the  gainer in the rest of the text), and an evolved G-type secondary (which we will refer to as the donor), that has likely filled its Roche lobe and is losing mass to its companion. Spectroscopic evidence indicates the presence of several layers of circumstellar matter in the system, and allows for the existence of a permanent accretion disk \citep{sahade82}. According to these characteristics, AU Mon belongs to the class of hot, massive Algols.

The orbit of the system is circular, with an orbital period of about 11 days. There is an additional periodicity in the total light of the system, with a period of about 417 days and an amplitude of about 0.25 mag. This variability was discovered by \citet{lorenzi80a} and interpreted by \citet{peters91, peters94} as resulting from periodic fluctuations in the mass transfer rate, which may be caused by oscillations of the donor star about its critical Roche surface.

A thorough overview of previous knowledge about the system can be found in the recent paper by \citet{des10}, who performed a very detailed analysis of the system based on CoRoT photometry (the same data this study is based on), a collection of previously published ground-based photometry, and on high-resolution spectroscopy. Based on the analysis of ground-based data, they concluded that the long-term variation originates from changes in the transparency of circumbinary material. They have also found periodicities shorter than the orbital period in CoRoT data. Two reliable frequencies of 10.4 and 8.3 $d^{-1}$ were tentatively assigned to the pulsation of the gainer, and the power excess in low-frequency region to the solar-like oscillations of the donor.

However, the attempt by \citet{des10} to model AU Mon as a semidetached system wasn't entirely successful. Most notably, the semidetached model could not reproduce the Rossiter-McLaughlin effect (\citealt{rossiter24,mc24}), and required an unusually large gravity darkening exponent to fit the light-curves. The authors therefore suggested that the system should be analyzed using a model which includes the gas stream and the accretion disk around the primary component.

There are several spectroscopic indications for the existence of an accretion disk in AU Mon, discussed in detail by \citet{des10}. The double-peaked $H_{\alpha}$ emission line, and the variation of $H_{\alpha}$ and $H_{\beta}$ line profiles with the orbital phase \citep{atwood10}, can be successfully explained with a model of the system with an accretion disc around the gainer. \citet{pp98} have found evidence of disk absorption in the $\rm{Fe_{\ III}}$ (UV1) resonance lines. Moreover, according to the criterion of \citet{LubowShu}, AU Mon should have a permanent accretion disk \citep{richards99}.

These spectroscopic indicators and the difficulties in reconciling the semidetached model with the observations, prompted us to repeat the light curve analysis of CoRoT photometry using a binary system model with an accretion disk. The model includes active regions on the disk edge, accounting for the gas stream and the spiral arms in the disk. We aim to demonstrate that the accretion disk model fits the CoRoT data better than the semidetached model.

\section{The model of the accretion disk}
\label{model}

The basic elements of the binary system model with a plane-parallel accretion disk and the corresponding light-curve synthesis procedure are described in detail by \citet {djura92, djura96}.

We assume that the disk in AU Mon is optically and geometrically thick. The disk edge is approximated by a cylindrical surface. In the current version of the model \citep{djura08}, the thickness of the disk can change linearly with radial distance, allowing the disk to take a conical shape (convex, concave or plane-parallel). The geometrical properties of the disk are determined by its radius ($R_d$), its thickness at the edge ($d_e$) and the thickness at the center ($d_c$). This way of approximating the shape of the accretion disk is justified by the current hydrodynamical modeling of mass transfer in close binary systems - see, e.g., \citet{bis00}, \citet{Harmanec}, \citet{nazar03, nazar06a, nazar06b}.

The cylindrical edge of the disk is characterized by its temperature, $T_d$, and the conical surface of the disk by a radial temperature profile obtained by modifying the temperature distribution proposed by \citet{zola91}:

\begin{equation}
T(r) = T_d + (T_\ast - T_d )\left[ {1 - \left({\frac{r - R_\ast }{R_d - R_\ast }} \right)} \right]^{a_T }
\label{eq1}
\end{equation}

We assume that the disk is in physical and thermal contact with the gainer, so the inner radius and temperature of the disk are equal to the temperature and radius of the star ($R_*$, $T_*$). The temperature of the disk at the edge ($T_d$) and the temperature exponent ($a_T$), as well as the radii of the star ($R_*$) and of the disk ($R_d$) are free parameters, determined by solving the inverse problem.

The model of the system is refined by introducing active regions on the edge of the accretion disk. The active regions have higher local temperatures so their inclusion results in a non-uniform distribution of radiation. The existence of such regions (hot and bright spots), can be explained by the gas dynamics of interacting binaries - see e. g. \citet{Heemskerk}, \citet{bis98, bis00, bis05}, \citet{Harmanec}, \citet{nazar03, nazar06a, nazar06b}. Based on these investigations of systems with accretion disks, we have recently updated our model to allow inclusion of up to three such active regions: a hot spot (hs) and two bright spots (bs1 and bs2). Including the active regions in the model leads to significant improvement of the quality of the fit.

The hot spot in our model is a rough approximation of the "hot line" which forms at the edge of the gas stream flowing from the donor to the disk. Due to the infall of an intensive gas-stream, the disk surface in the region of the hot spot becomes deformed as the material accumulates at the point of impact, producing a local deviation of radiation from the uniform azimuthal distribution \citep[for more details, see][]{djura08}. In the model, this deviation is described by the angle ${\rm \theta_{rad}}$ between the line perpendicular to the local disk edge surface and the direction of the hot spot's maximum radiation. Depending on ${\rm \theta_{rad}}$, the maximum of the hot spot flux can be slightly shifted in orbital phase, changing the light-curve asymmetry around the secondary maximum and in the region of the primary minimum.

The bright spots in the model approximate the spiral structure of the disk, predicted by hydrodynamical calculations \citep{Heemskerk}. The tidal force exerted by the donor star causes a spiral shock, producing one or two extended spiral arms in the outer part of the disk. The bright spots can also be interpreted as regions where the disk significantly deviates from the circular shape.

\section{ The light-curve analysis}

This study is based on photometric measurements of AU Mon collected by the CoRoT space mission. The data consist of 139704 individual measurements with the time sampling of 32 seconds, taken during 56 days, encompassing five full orbital periods of the system.

The observations were folded to orbital period using the ephemeris from \citet{des10}:

\begin{equation}
T_{min1}=\rm{HJD}2454136.6734(2) + 11.^d 1130374(1) \times E
\label{eq2}
\end{equation}

Since in this study we were mainly interested in the physical and orbital characteristics of the system, we downsampled and divided the data into five light curves, each consisting of 1000 points. The light level of the system slowly decreases over the time span of the observations, hence we normalized each light curve to the maximum of the first one at phase 0.25 (the total brightness of the system). In Fig.~\ref{fAUMon} (top five panels), this maximum light level is shown by the horizontal dashed line. The normalized light curves were analyzed separately.

The light-curve analysis was performed using the inverse-problem solving method \citep{djura92} based on the simplex algorithm, and the above described model of a binary system with an accretion disk.

To obtain reliable estimates of the system parameters, a good practice is to restrict the number of free parameters by fixing them to values obtained from independent sources. Thus we fixed the spectroscopic mass ratio to $q=M_c/M_h=0.17$ (where the subscripts $h$ and $c$ refer to the hotter gainer and the cooler donor), and the temperature of the donor to $T_c=5750K$, based on the spectroscopic study by \citet{des10}. The temperature of the gainer was a free parameter in the preliminary analysis, and for the final solution we adopted the average of the individual solutions for the five successive CoRoT light curves. The average value of $T_h=15870K$, obtained in this way, was then kept fixed in the final analysis. We adopted the factor of non-synchronous rotation of the gainer $f_h=5.2$ from the study by \citet{Glazunova08}, and considered the rotation of the donor as synchronous ($f_c=1.0$), assuming that it has filled its Roche lobe (i. e. the filling factor of the donor was set to $F_c=1.0$). In addition, we set the gravity darkening coefficient and the albedo of the gainer to $\beta_h=0.25$ and $A_h=1.0$ in accordance with von Zeipel's law for radiative shells and complete re-radiation; for the donor we set $\beta_c=0.08$ and $A_c=0.5$, as is appropriate for stars with convective envelopes according to \citet{lucy}, \citet{rucinski} and \citet{rafert}.

The limb-darkening for the components was taken into account by using the non-linear approximation by \citet{claret}, which can be expressed as 

$$
\frac{I(\mu)}{I(1)}=1-a_1(1-\mu^{1/2})-a_2(1-\mu)-a_3(1-\mu^{3/2})-a_4(1-\mu^2)
$$

\noindent where $a_{1,2,3,4}$ are the limb-darkening passband-specific coefficients, $I(1)$ is the passband-specific intensity at the center of the stellar disc, and $\mu=\cos \gamma$, where $\gamma$ is the angle between the line of sight and the local surface normal. The central intensity is calculated for the effective wavelengths of the observations (600 nm for the CoRoT light curves and 550 nm for the V-band light curve), using a simple blackbody approximation. For a given metallicity, the values of the passband-specific limb-darkening coefficients are derived from the current values of the stellar effective temperature ${\rm T_{eff}}$ and surface gravity ${\rm log} \ g$ in each iteration, by {\it bi-linear} interpolation \citep{press}, of both quantities from tables of \citet{claret} for the V light curves and of \citet{sing} for the CoRoT light curves. The procedure is described in more detail by \citet{djura04}. The limb-darkening was applied to the disk in the same way, with $\log g$ corresponding to the middle of the disk radius.

\begin{table*}
\caption{Results of the analysis of CoRoT and V-band light-curves, obtained by solving the inverse problem for the Roche model with an accretion disk around the more-massive (hotter) component, and with the simple semidetached Roche model}.
\begin{minipage}{200mm}
\label{TabAUMon}
\[
\begin{array}{@{\extracolsep{+0.0mm}}lllllllll@{}}
\hline
\noalign{\smallskip}
{\rm Quantity}  &{\rm CoRoT \ (1)} & {\rm CoRoT \ (2)} & {\rm CoRoT \ (3)} & {\rm CoRoT \ (4)} & {\rm CoRoT \ (5)} & {\rm MEAN_{CoRoT}} & {\rm MEAN_{CoRoT}} & {\rm V-filter}  \\
& {\rm Disk \ model} & {\rm Disk \ model}  & {\rm Disk \ model}  & {\rm Disk \ model}  & {\rm Disk \ model}  & {\rm Disk \ model}   & {\rm Roche \ model}  & {\rm Disk \ model}\\
\noalign{\smallskip}
\hline
\noalign{\smallskip}
n                   & 1000      & 1000      & 1000      & 1000      & 1000      &      & & 2682    \\
{\rm \Sigma(O-C)^2} & 0.0497    & 0.0410    & 0.0636    & 0.0480    & 0.0454    &      & & 2.6880  \\
{\rm \sigma_{rms}}  & 0.0070    & 0.0064    & 0.0080    & 0.0069    & 0.0067    &      & & 0.0317  \\
q                   & 0.17      & 0.17      & 0.17      & 0.17      & 0.17      & 0.17 \pm 0.03  & 0.17 \pm 0.03& 0.17  \\
i {\rm [^{\circ}]}  & 80.07     & 80.09     & 80.06     & 80.09     & 80.07     & 80.1 \pm 0.6   & 78.8 \pm 0.6 & 80.08 \\
{\rm F_d}           & 0.575     & 0.578     & 0.532     & 0.560     & 0.518     & 0.55 \pm 0.03  &  & 0.575 \\
{\rm T_d} [{\rm K}] &  5140     &  5174     &  5209     & 5256      &  5192     & 5190 \pm 70    &  & 5250 \\
{\rm d_e} [a_{\rm orb}] & 0.040 & 0.039     & 0.039     & 0.039     & 0.039     &0.040 \pm 0.005 &  & 0.042 \\
{\rm d_c} [a_{\rm orb}] & 0.010 & 0.010     & 0.010     & 0.010     & 0.010     & 0.01           &  & 0.01 \\
{\rm a_T}               & 6.50  & 6.50      & 6.50      & 6.50      & 6.50      & 6.5            &  & 6.5  \\
{\rm F_h}               & 0.565 & 0.565     & 0.565     & 0.565     & 0.565     & 0.565 \pm 0.002& 0.62 \pm 0.01 & 0.565 \\
{\rm T_h} [{\rm K}]     & 15894 & 15905     & 15886     & 15902     & 15890     & 15890 \pm 400  & 14100 \pm 400 & 15890 \\
{\rm T_c} [{\rm K}]     & 5750  &  5750     &  5750     &  5750     &  5750     & 5750           & 5750  & 5750 \\
{\rm A_{hs}=T_{hs}/T_d}             &  1.68     &  1.68     &  1.70     &  1.70     &  1.68      &  1.7 \pm 0.1 & & 1.68 \\
{\rm \theta_{hs}}{\rm [^{\circ}]}   &  19.8     &  19.5     &  19.7     &  19.9     &  19.6      & 19.7 \pm 0.4 & & 20.0 \\
{\rm \lambda_{hs}}{\rm [^{\circ}]}  & 325.6     & 329.0     & 330.7     & 334.2     & 337.6      & 331  \pm 6   & & 312.4\\
{\rm \theta_{rad}}{\rm [^{\circ}]}  & -33.2     & -34.8     & -34.6     & -34.7     & -32.1      & -34  \pm 7   & & -32.1\\
{\rm A_{bs1}=T_{bs1}/T_d}           &  1.34     &  1.47     &  1.35     & 1.48      &  1.51      & 1.4  \pm 0.2 & & 1.31 \\
{\rm \theta_{bs1}} {\rm [^{\circ}]} &  39.4     &  39.6     &  39.3     &  30.7     &  17.0      & 33   \pm 12  & & 36.5 \\
{\rm \lambda_{bs1}}{\rm [^{\circ}]} & 177.1     & 178.5     & 178.4     & 164.5     & 158.0      & 171  \pm 13  & & 117.9\\
{\rm A_{bs2}=T_{bs2}/T_d}           &  1.39     &  1.41     & 1.28      &  1.33     &  1.25      & 1.3  \pm 0.1 & & 1.50 \\
{\rm \theta_{bs2}} {\rm [^{\circ}]} &  49.7     &  49.4     &  49.2     &  49.8     &  48.5      & 49   \pm 6   & & 50.0 \\
{\rm \lambda_{bs2}}{\rm [^{\circ}]} &  60.0     &  47.2     &  52.6     &  51.0     &  53.8      & 53   \pm 14  & & 73.7 \\
{\rm \Omega_h}                      & 8.614     & 8.613     & 8.611     & 8.614     & 8.616     &  8.6   \pm 0.5 & 7.9 \pm 0.6  & 8.612 \\
{\rm \Omega_c}                      & 2.156     & 2.156     & 2.156     & 2.156     & 2.156     &  2.2   \pm 0.2 & 2.16\pm 0.05 & 2.156 \\
\noalign{\smallskip} \hline \noalign{\smallskip}
\cal M_{\rm_h} {[\cal M_{\odot}]} & 6.957     & 6.957     & 6.957     & 6.957     & 6.957     & 7.0  \pm 0.3  & 7.0 \pm 0.4 & 6.957  \\
\cal M_{\rm_c} {[\cal M_{\odot}]} & 1.183     & 1.182     & 1.183     & 1.182     & 1.182     & 1.20 \pm 0.2  & 1.2 \pm 0.2 & 1.183  \\
\cal R_{\rm_h} {\rm [R_{\odot}]}  & 5.081     & 5.082     & 5.083     & 5.081     & 5.080     & 5.1  \pm 0.5  & 5.6 \pm 0.5 & 5.082  \\
\cal R_{\rm_c} {\rm [R_{\odot}]}  &10.083     &10.083     &10.083     &10.083     &10.083     & 10.1 \pm 0.5  & 10.1\pm 0.5 & 10.083 \\
{\rm log} \ g_{\rm_h}             & 3.868     & 3.868     & 3.868     & 3.868     & 3.869     & 3.87 \pm 0.05 & 3.78\pm 0.06& 3.868  \\
{\rm log} \ g_{\rm_c}             & 2.504     & 2.504     & 2.504     & 2.503     & 2.503     & 2.50 \pm 0.02 & 2.50\pm 0.02& 2.504  \\
M^{\rm h}_{\rm bol}               &-3.138     &-3.141     &-3.136     &-3.140     &-3.136     & -3.1 \pm 0.4  &-2.8 \pm 0.4 & -3.137 \\
M^{\rm c}_{\rm bol}               &-0.210     &-0.210     &-0.210     &-0.210     &-0.210     &-0.21 \pm 0.09 &-0.21\pm 0.09& -0.210 \\
a_{\rm orb}  {\rm [R_{\odot}]}    & 42.11     & 42.11     & 42.11     & 42.10     & 42.10     & 42.1 \pm 0.4  & 42.1\pm 0.5 & 42.1   \\
\cal{R}_{\rm d} {\rm [R_{\odot}]} & 13.20     & 13.28     & 12.23     & 12.87     & 11.90     & 12.7 \pm 0.6  & & 13.2   \\
\rm{d_e}  {\rm [R_{\odot}]}       &  1.69     &  1.66     & 1.65      &  1.66     &  1.66     & 1.66 \pm 0.05 & & 1.78   \\
\rm{d_c}  {\rm [R_{\odot}]}       &  0.42     &  0.42     &  0.42     &  0.42     &  0.42     & 0.42 \pm 0.01 & & 0.42   \\
\noalign{\smallskip} \hline \end{array} \] \bigskip \end{minipage}

\begin{minipage}{\textwidth}
\footnotesize
FIXED PARAMETERS: $q={\cal M}_{\rm c}/{\cal M}_{\rm h}=0.17$ - mass ratio of
the components, ${\rm T_h=15890 K}$  - temperature of the more-massive (hotter)
gainer, ${\rm T_c}=5750 K$ - temperature of the less-massive (cooler) donor,
${\rm F_c}=1.0$ - filling factor for the critical Roche lobe of the donor,
$f{\rm _h}=5.2 ; f{\rm _c}=1.00$ - non-synchronous rotation coefficients of
the gainer and donor respectively, ${\rm \beta_h=0.25 ; \beta_c=0.08}$ -
gravity-darkening coefficients of the gainer and donor, ${\rm A_h=1.0 ;
A_c=0.5}$  - albedo coefficients of the gainer and donor.

\smallskip \noindent QUANTITIES: $n$ - number of observations,
${\rm \Sigma (O-C)^2}$ - final sum of squares of residuals between observed
(LCO) and synthetic (LCF) light-curves, ${\rm \sigma_{rms}}$ - root-mean-square
of the residuals, $i$ - orbit inclination (in arc degrees),
${\rm F_d=R_d/R_{yc}}$ - disk dimension factor (ratio of the disk radius to
the critical Roche lobe radius along y-axis), ${\rm T_d}$ - disk-edge
temperature, $\rm{d_e}$, $\rm{d_c}$,  - disk thicknesses (at the edge and at
the center of the disk, respectively) in the units of the distance between the
components, $a_{\rm T}$ - disk temperature distribution coefficient,
${\rm F_h}=R_h/R_{zc}$ - filling factor for the critical Roche lobe of the
hotter, more-massive gainer (ratio of the stellar polar radius to the
critical Roche lobe radius along z-axis),
${\rm A_{hs,bs1,bs2}=T_{hs,bs1,bs2}/T_d}$ - hot and bright spots' temperature
coefficients, ${\rm \theta_{hs,bs1,bs2}}$ and ${\rm \lambda_{hs,bs1,bs2}}$ -
spots' angular dimensions and longitudes (in arc degrees), ${\rm \theta_{rad}}$
- angle between the line perpendicular to the local disk edge surface and the
direction of the hot-spot maximum radiation, ${\rm \Omega_{h,c}}$ -
dimensionless surface potentials of the hotter and cooler components,
$\cal M_{\rm_{h,c}} {[\cal M_{\odot}]}$, $\cal R_{\rm_{h,c}} {\rm [R_{\odot}]}$
- stellar masses and mean radii of stars in solar units,
${\rm log} \ g_{\rm_{h,c}}$ - logarithm (base 10) of the system components
effective gravity, $M^{\rm {h,c}}_{\rm bol}$ - absolute stellar bolometric
magnitudes, $a_{\rm orb}$ ${\rm [R_{\odot}]}$, $\cal{R}_{\rm d} {\rm [R_{\odot}]}$,
$\rm{d_e} {\rm [R_{\odot}]}$, $\rm{d_c} {\rm [R_{\odot}]}$ - orbital semi-major axis,
disk radius and disk thicknesses at its edge and center, respectively, given in the
solar radius units.
\vspace{0.5cm}
\end{minipage}
\end{table*}

\section{Results}
\label{results}

The results of the light-curve analysis based on the described model of AU Mon are given in Table~\ref{TabAUMon}. The first column contains parameter designations, and the following five columns give the values derived from each of the five CoRoT light-curves, with the mean values and their estimated uncertainties. The uncertainties were estimated from a set of solutions obtained for the five observed light curves and three different values of the mass ratio: $q_1=0.14$, $q_2=0.17$ and $q_3=0.20$ (chosen according to the error assigned to the mass ratio by \citealt{des10}, $q=0.17\pm 0.03$), resulting in a total of 15 values for each parameter. The uncertainties given in Table~\ref{TabAUMon} are the maximal deviations of these values from the mean. Table~\ref{TabAUMon} also lists the results of applying a simple Roche model \citep[see e.g.][]{djura92} to the CoRoT light curves and the results of applying the accretion disk model to the ground-based V-band light curves (discussed in detail in Section~\ref{longterm}). The first three rows of Table~\ref{TabAUMon} present the number of points in the light curve ($n$), the final sum of the squares of the residuals between the observed (LCO) and the synthetic (LCF) light curves, $\sum (O-C)^2$, and the root-mean-square of the residuals $\sigma_{rms}$, respectively.

\begin{figure*}
\centering
\begin{minipage}{\textwidth}
\centering
\includegraphics[width=12.5cm,angle=-90]{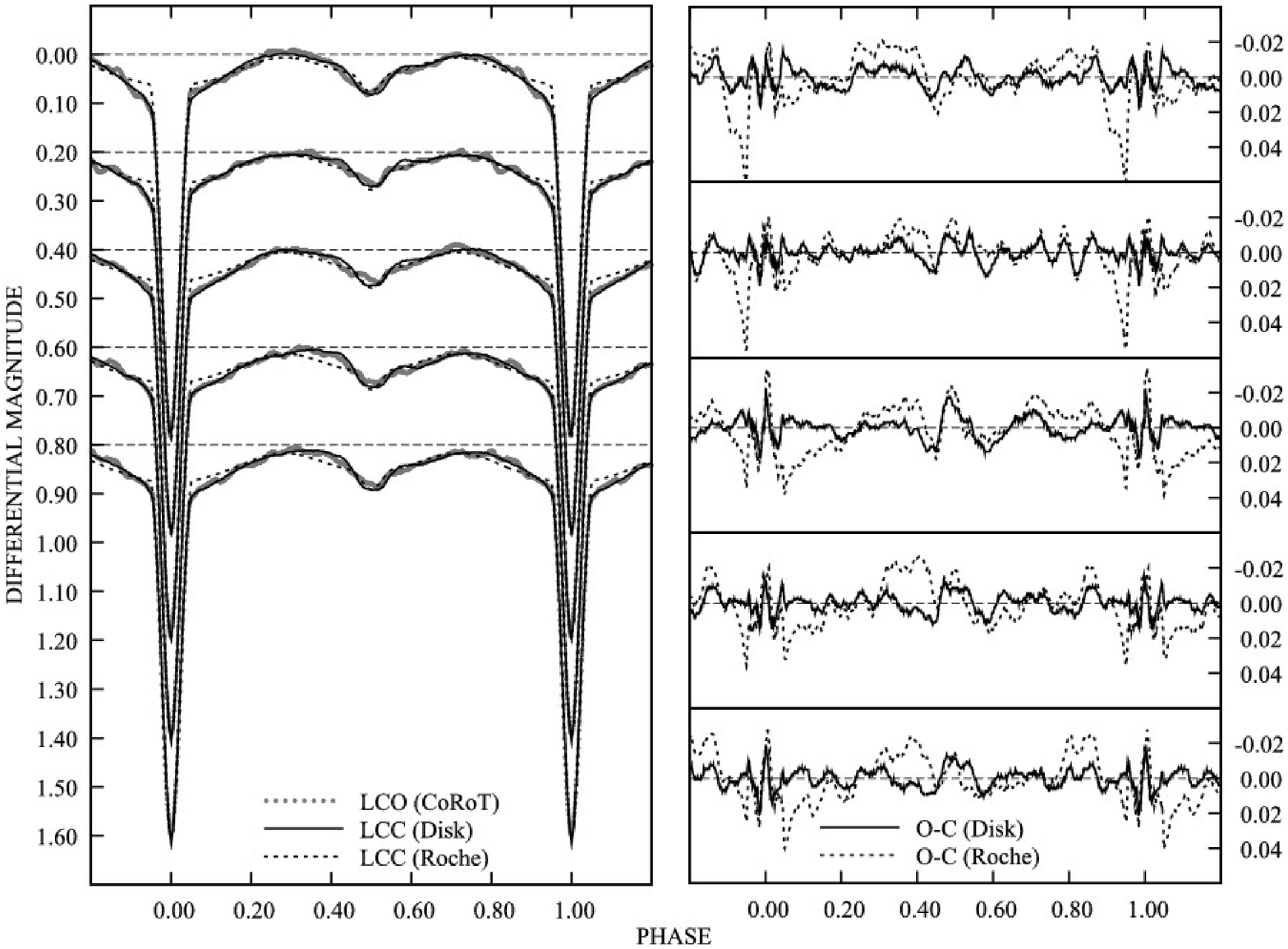}
\caption{Left: CoRoT observed (LCO) and synthetic (LCC) light-curves of AU Mon (the
normalized maximum light level is shown by the horizontal dashed line) obtained by applying
a simple semidetached Roche model (dashed line) and an accretion disk model (solid line) with
the active regions on the disk edge. The successive individual light curves are shifted by
0.2 magnitude; Right: the final O-C residuals between the CoRoT observed (LCO) and optimal
synthetic (LCC) light curve, obtained by applying the simple semidetached Roche model (Roche)
and accretion disk model (Disk).}
\label{fAUMon}
\end{minipage}
\end{figure*}

The best fit model of AU Mon contains an optically and geometrically thick accretion disc around the hotter, more massive gainer star. With a radius of $R_d\approx13 R_{\odot}$, the disk is more than twice as large as the central star ($R_h\approx5 R_{\odot}$). The disk has a moderately concave shape, with central thickness of $d_c\approx 0.4 R_{\odot}$ and the thickness at the edge of $d_e\approx1.6 R_{\odot}$. The temperature of the disk increases from {\bf $T_d=5190 K$} at its edge, to $T_h=15870 K$ at the inner radius (where it is in thermal and physical contact with the gainer), according to Eq.~\ref{eq1}, with the temperature profile exponent $a_T=6.5$. The effective temperature of the disk is significantly higher than the temperature at its edge.

\begin{figure*}
\centering
\begin{minipage}{\textwidth}
\centering
\includegraphics[]{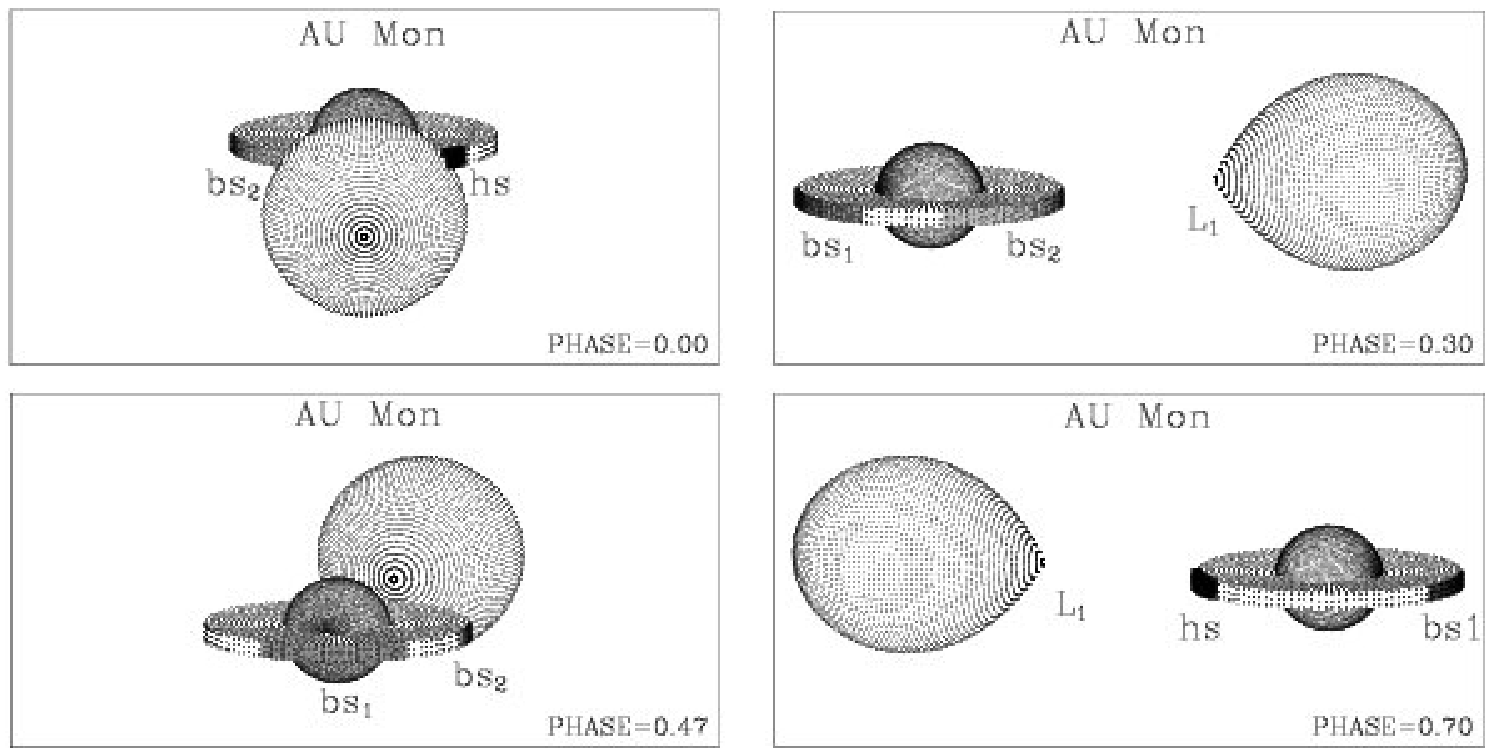}
\caption{The views of the optimal accretion disk model of
the system AU Mon at orbital phases 0.00, 0.30, 0.47 and 0.70, obtained with
parameters estimated by the CoRoT light curve analysis.}
\label{viewAUMon}
\end{minipage}
\end{figure*}

We were able to model the asymmetry of the light curve very precisely by incorporating three regions of enhanced radiation on the accretion disk: the hot spot (hs), and two bright spots (bs1 and bs2). The hot spot (hs) is situated at longitude $\lambda_{hs}\approx 330^\circ$, roughly between the components of the system, at the place where the gas stream falls onto the disk. The longitude $\lambda$ is measured clockwise (as viewed from the  direction of the +Z-axis, which is orthogonal to the orbital plane) with respect to the line connecting the star centers (+X-axis), in the range $0^\circ-360^\circ$. The temperature of the hot spot is approximately 70\% higher then the disk edge temperature, i.e. $T_{hs}\approx 8700 K$. The hot spot can be interpreted as a rough approximation of the "hot line" which forms at the edge of the gas stream between the components. According to \citet{atwood10}, a gas-stream with a temperature near $8000 K$ can explain the excess red- and blue-shifted $H_\alpha$ emission near phases 0.2 and 0.7, respectively. We note that this spectroscopic result is an independent confirmation of our photometrically estimated temperature of the gas stream.

Although including the hot spot region into the model significantly improves the fit, it cannot explain the light-curve asymmetry completely. By introducing two additional bright spots (bs1 and bs2), larger than the hot spot and located on the disk edge at $\lambda_{bs1}\approx 170^\circ$ and $\lambda_{bs2}\approx 50^\circ$, the fit becomes much better.

The bright spots can be related to the spiral shocks that result from radiative cooling and form at the outer boundary of the disk. Since the disk is large, filling about 56\% of the gainer's critical Roche surface, the tidal forces exerted by the donor can cause a spiral-shaped tidal shock in the disk - see e.g. \citet{Heemskerk}. Such a shock wave can produce one or two extended spiral arms in the outer parts of the disk.

The first arm, represented in our model by a bright spot (bs1), is located on the disk edge at longitude $\lambda_{bs1}\approx 170^\circ$. This is also a region where we can expect loss of matter from the gas stream and the disk through the Lagrangian point ${\rm L_3}$, forming some kind of a circumbinary shell.

The fit was additionally improved by introducing the second bright spot (bs2), located at longitude $\lambda_{bs2}\approx 50^\circ$. This spot is the largest active region, with a temperature about 30\% higher than the disk edge temperature. It can be interpreted as the second spiral arm in the disk.

We note that the system can also be modeled with active regions (dark spots) on the donor, and without the active regions on the accretion disk. Such a model would explain the period-to-period variations in the light curves by the presence, development and migration of spots over the surface of the donor. However, the model with active regions on the accretion disk seems to be more appropriate. Namely, the relatively fast variations of the light curves are more likely to originate from the changes in the disk structure, produced by variable mass outflow from the donor, than from the motion of stellar spots, not expected to take place on these timescales.

In order to make a comparison between the results of our study and that of \citet{des10}, we made several trial runs with a simple semidetached Roche model of AU Mon. The semidetached model cannot fit the observations as well as the model with an accretion disk. The fit can be improved by using an anomalously large gravity darkening exponent of the donor, as done by \citet{des10}. Our previous investigations of anomalously high values of gravity darkening exponents reported in literature (\citealt{djura03,djura06}) have shown that such anomalies are usually the consequence of an inadequate model used for the light curve analysis. The parameters we obtained for a semidetached Roche model (without the disk), with the theoretical value of gravity darkening are shown in Table~\ref{TabAUMon} (column 7). The semidetached model gives a slightly smaller inclination  (by approximately 2\%), and a larger radius of the gainer (by approximately 9\%), with a lower temperature (by approximately 10\%), which was to be expected since the large accretion disk, surrounding the gainer, is characterized by a lower effective temperature. In summary, the absence of the accretion disk in the semidetached model is compensated by a smaller inclination, and a larger, cooler gainer. Since the spectral type of the gainer was reported to be B5 (\citealp[see e.g.][and the references within]{des10}), the gainer's temperature estimated with the accretion disk model is in better agreement with this spectral type than the temperature estimated with the simple semidetached model.

Another problem with the semidetached model used by \citet{des10} is its inability to reproduce the Rossiter-McLaughlin (RM) effect (\citealt{rossiter24,mc24}). To improve the quality of the radial velocity curve fit to the observations, \citet{des10} had to enlarge the radius of the donor much over the value obtained from the light curve analysis (see Figure 11 of their paper). This discrepancy can also be solved by introducing an accretion disk around the gainer. Namely, from the bottom panel of Fig.~\ref{fAUMonV}, which shows the system at orbital phase 0.5 (in the secondary eclipse), it is intuitively clear that the accretion disk contributes to the RM effect in a si\-mi\-lar way as the enlargement of the donor. This remains to be proven by spectroscopic modeling. Nevertheless, our qualitative conclusion is that the enlargement of the donor is unnecessary if the system is modeled with an accretion disk.

\begin{figure}
\centering
\includegraphics[]{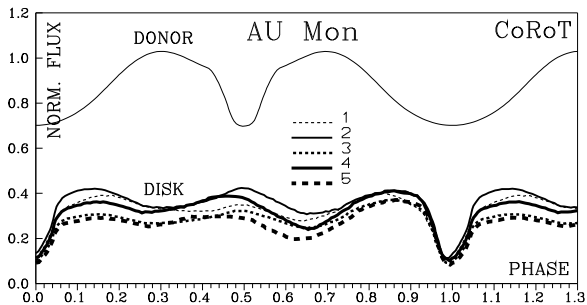}
\caption{The calculated fluxes of the donor and of the accretion disk, normalized to the donor flux at the phase 0.25 and given for all five CoRoT light curves.}
\label{fluxAUMon}
\end{figure}

The fits are shown in the left panel of Fig.~\ref{fAUMon} for each orbital period. The successive individual light curves are shifted by 0.2 mag to save space. The figure shows the observations (LCO - CoRoT) and two fits: the fit of the described accretion disk model (LCC - Disk) and {\bf the fit} of a simple semidetached Roche model (LCC - Roche) with the fixed gravity darkening exponent. The residuals for both models are given in the right panel of Fig.~\ref{fAUMon}. The model with the accretion disk, made with the theoretically expected gravity darkening exponent $\beta=0.08$ for the convective donor, evidently provides a superior fit to the observations.

Figure~\ref{viewAUMon} shows the appearance of AU Mon in orbital phases 0.00, 0.30, 0.47 and 0.70. The phases were chosen so as to make both components and the active regions on the disk edge visible.

Finally, Fig.~\ref{fluxAUMon} presents the individual synthetic fluxes of the donor and the disk, normalized to donor brightness at orbital phase 0.25. The main contribution to the flux comes from the hotter, more massive gainer, and is approximately 3 times greater than the contribution from the donor, so it is not shown here to save space. In the present model, the donor and the gainer fluxes are the same for all five orbital periods in CoRoT data, and the shape variations of the light curves are entirely due to the changes in the flux emitted by the accretion disk, probably produced by the variable mass outflow from the donor.

\subsection{The long-term variability}
\label{longterm}

The long-term variability in total brightness of the system was first explained as the result of fluctuating mass transfer rate by \citet{peters91, peters94}. The rate of mass transfer may change periodically due to the oscillations of the donor star around its critical Roche lobe, causing changes in the size and density of the accretion disk and/or circumbinary matter. \citet{vivek98}, who studied the photoelectric measurements of \citet{lorenzi80b}, using a similar semidetached model as \citet{des10}, suggested that the long-term variation may be due to the precession of the accretion disk around the gainer, caused by the gravitational perturbations by the donor.

The ephemeris of the long-term variability calculated by \citet{des10} is:

\begin{equation}
T_{max}=\rm{HJD}2443105.1(\pm1.4) + 416.^d 9(\pm8.7) \times E.
\label{eq3}
\end{equation}

This puts the CoRoT observations at the minimum of the total light, between phases 0.46 and 0.59 of the long-term variation.

In order to study the behavior of the model with accretion disk during the entire long cycle, we used the ground based photometric data from \citet{lorenzi80b,lorenzi85} and \citet{kilkenny85} in the V passband (courtesy of M. Desmet).

\begin{figure}
\centering
\includegraphics[]{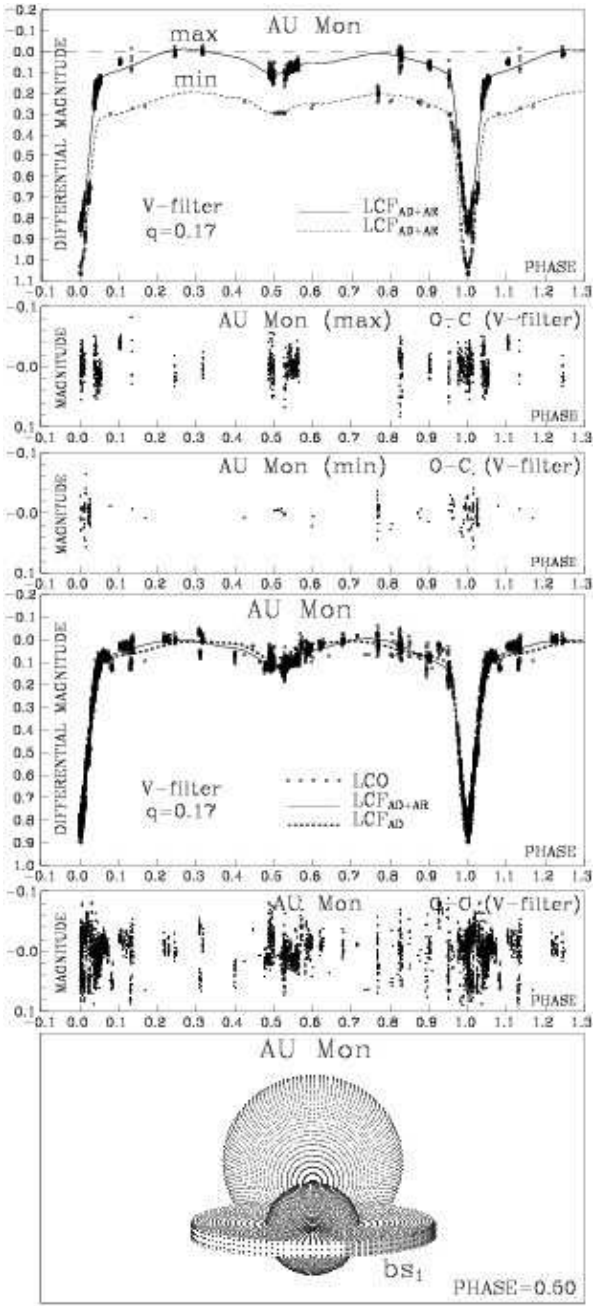}
\caption {From top to bottom: V-band observed light curves (LCO) at maximum (max) and minimum (min) of the long-term period (${\rm P_{long}\sim 417 \ days}$) and corresponding synthetic (LCF) light-curves of AU Mon obtained by applying  accretion disk model with active regions (AD+AR) - the normalized maximum light level is shown by the dashed line; The final O-C residuals for the maximum and minimum light curves obtained by applying accretion disk model with active regions (AD+AR); Observed (LCO) V-band light curve (after removal of the long-term period) and the corresponding fitting curves (${\rm LCF_{AD+AR},LCF_{AD}}$) obtained by applying accretion disk model with (AD+AR) and without (AD) active regions; the corresponding O-C residuals  obtained for the accretion disk model with active regions (AD+AR); the view of the model of the system AU Mon at orbital phase 0.50 obtained with parameters estimated from the V-band light curve analysis.}
\label{fAUMonV}
\end{figure}

After folding the light curves to the long period according to the ephemeris given by Eq.~\ref{eq3}, we attempted to model the data from the maximum (phases 0.9 to 1.1) and from the minimum (phases 0.4 to 0.6) of the long period, using the model obtained with CoRoT data, with the quantities related to the accretion disk as free parameters. We found that the variation in total brightness cannot be explained by changes in the size of the accretion disk and the sizes and locations of the active regions on the disk edge. Such changes would cause variations in depth of the light curve minima, but there is no evidence that the amplitude of the light curve is changing over the long period. Our analysis thus confirms the conclusion of \citet{des10}, that the long-term variation must be the result of variable attenuation of the total light of the system by some form of a circumbinary shell.

The accretion disk model, when used with the same methods and free parameters we employed in the analysis of the CoRoT data, fits the ground-based data very well and the parameters obtained in this way (given in column 8 of Table~\ref{TabAUMon}) are in good agreement with those obtained from the CoRoT light curves. We conclude that our model describes both data sets adequately.

The ground-based data and the synthetic light-curves from our model are shown in Figure~\ref{fAUMonV}. The top panel shows the fit to the light curves in the minimum and maximum of the long cycle, the middle panel shows the fit to the entire ground-based dataset, reduced for the effects of the long-term variation, and the bottom panel gives a representation of the model obtained from the entire ground-based dataset in the phase of the secondary eclipse.

\subsection{The short-term variability}
\label{shortterm}

There are evident short-term variations in the CoRoT light curves of AU Mon. The residuals after subtracting the synthetic light curve from the observations (Figure~\ref{fAUMon}, right panel), also show quasi-periodic variability. Two features stand out: a lump in the secondary minimum (around phase 0.5) which is present to some degree in all five residual curves and is partially due to the inability of the model to fit the observations perfectly. However, the shape and amplitude of the lump change from period to period, suggesting the presence of intrinsic variability in either the G type secondary or in the accretion disk or both.

The other notable feature in the residual light curves is the fast and seemingly periodic variation inside the primary minimum (around phase 1.0). \citet{des10}, who also noted the fast variation, suggested that its origin is a non-uniform brightness distribution of the accretion disk.

There is, however, an alternative interpretation: we believe that the fast variation during the primary minimum of AU Mon is caused by a hidden mode of oscillation of the primary \citep[see e. g.][]{mkrt05}. The origin of the variation is unlikely to be the secondary star or the accretion disk, because these components are at least partially visible during the entire orbital cycle, while the fast variation occurs only in the primary minimum. If our interpretation is correct, it also represents an indication that the primary star is indeed producing oscillations.

\begin{figure}
\centering
\includegraphics[width=8.5cm]{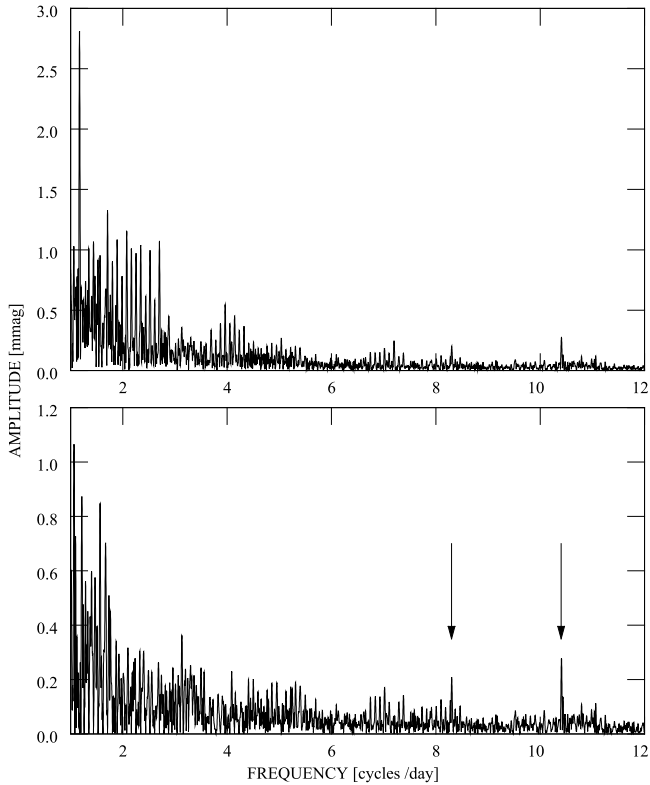}
\caption{The Fourier periodogram computed from the $O-C$ residuals in the frequency range from 0 to 12 $d^{-1}$, before (upper panel) and after (lower panel) prewhitening the data for the harmonics of the orbital frequency.}
\label{periodogram}
\end{figure}

We performed a frequency analysis of the residuals remaining after fitting our model to the observations using \textsc{Period04}, a signal analysis program based on Fourier Transform methods \citep{lenz05}. A time series of 5000 points was formed by concatenating the residuals remaining after fitting each orbital period, allowing the computation of a periodogram in the frequency range from 0 to 45 $d^{-1}$ with a frequency resolution of 0.0001 $d^{-1}$, shown in part in the upper panel of  Figure~\ref{periodogram}.

The low-frequency range is dominated by the harmonics of the orbital period. \citet{des10} were faced with the same problem during the frequency analysis of their residual data, and dealt with it first by cutting out the primary minima from their data set, and then by dividing the data with a spline-fit through the highest peaks in the PDM periodogram, with an effect of a high-pass filter.

Our attempt to remove the harmonics of the orbital frequency by cutting out regions around the primary and secondary minima did not result in a significant improvement in the appearance of the periodogram, so we used the method of iterative sine fitting (prewhitening) to remove about 40 harmonics from the signal. The resulting, cleaned periodogram is shown in the lower panel of Figure~\ref{periodogram}. The arrows indicate the two stable frequencies reported by \citet{des10}, at 10.40 and 8.30 $d^{-1}$. The frequencies and amplitudes of these two peaks, as determined from our residual data (listed in Table~\ref{frequencies}), are in good agreement with their results, indicating that in this case the frequency analysis is not very sensitive to the model of the binary system.

\begin{table}
\caption{Frequencies found in the $O-C$ residuals.}
\label{frequencies}
\begin{tabular}{lrrrrr}
\hline
ID & \multicolumn{2}{r}{Frequency} & Amplitude & SNR \\
 & $d^{-1}$ & $\mu Hz$ & $mmag$ & \\
\hline \hline
$f_{1}$ & 10.4079 & 120.46 & 0.28 & 7.5 \\
$f_{2}$ & 8.3036 & 96.11 & 0.21 & 5.3 \\
\hline
\end{tabular}
\end{table}

\section{Summary}

The motivation for this work came from the recent spectroscopic studies that offer compelling evidence of the presence of an accretion disk in the interacting eclipsing binary AU Mon. We analyzed the CoRoT photometric observations of AU Mon using a model that includes a geometrically and optically thick accretion disk. The model fits the observations very well. We note that we also modeled the system using a simple Roche model without an accretion disk, and found the final sum of the squares of the residuals between the observed and synthetic light-curves for this model to be significantly larger than for the model with the disk.

The accretion disk in our model is moderately concave and extends to roughly 60\% of the critical Roche lobe of the gainer. The asymmetry of the light curve is successfully modeled by incorporating three active regions on the accretion disk: a hot spot (hs), and two bright spots (bs1 and bs2), which in turn represent the "hot line" and two spiral arms of the disk and/or the deviations of the disk from a circular shape. The temperature of the hot spot ($T_{hs}\approx 8700 K$) is in good agreement with the gas-stream temperature $T\approx 8000 K$ obtained from $H_\alpha$ emission line profile modeling near orbital phases 0.2 and 0.7 \citep{atwood10}, which is an independent confirmation of our photometrically estimated temperature of the hot line region. \citet{atwood10} found that a disk with a maximum temperature of $T_{max}\approx 13700 K$ is in general agreement with the observed UV spectrum. The estimated temperature of our disk is in the range from 5100 K (at the edge of the disk) to 15870 K (in the inner part that is in contact with the gainer). The radial temperature profile of the disk, given by  Eq.~\ref{eq1}, shows that the maximum temperature of the disk estimated by \citet{atwood10} lies well within the range covered by our model.

The period-to-period variability of the CoRoT light-curves can be explained by the changes in the contribution of the disk flux to the total light, due to variations in the outflow from the donor. However, the long-term variability cannot be explained by the changes in the disk parameters alone, because such changes would cause variations in the amplitude of the light curve over the long period. As the changes of the light curve amplitude were not observed, the long-term variability of the system can only be interpreted as a result of a periodic change in the attenuation of the total light by an inhomogeneous circumbinary shell.

\section*{Acknowledgments}

The authors gratefully acknowledge Janos Nuspl and Mercedes Richards for the critical remarks and valuable suggestions, and M. Desmet for making the collection of ground based photometry of AU Mon available. The authors would also like to thank the anonymous referee for constructive comments, and O. Atanac\-kovi\'c and Z. Kne\v zevi\'c for helping us prepare the paper. We acknowledge the CoRoT space mission team, whose efforts have made this work possible. This research was funded by the Ministry of Science and Technological Development of the Republic of Serbia through the project "Stellar and Solar Physics" (No. 146003). The authors acknowledge the use of the electronic database SIMBAD operated by the CDS, Strasbourg, France, and the electronic bibliography maintained by the NASA/ADS system.

\bsp

\label{lastpage}

\end{document}